\documentclass[epj]{svjour}
\usepackage{graphics}
\usepackage{epsfig}
\newcommand{\boma}[1]{\mbox{\boldmath$#1$\unboldmath}}
\newcommand{\beq}{\begin{equation}}
\newcommand{\eeq}{\end{equation}}
\newcommand{\beqn}{\begin{displaymath}}
\newcommand{\eeqn}{\end{displaymath}}
\newcommand{\beqa}{\begin{eqnarray}}
\newcommand{\eeqa}{\end{eqnarray}}

  \newcommand{\Abar}{\bar{A}}
 \newcommand{\Bbar}{\bar{B}}
 \newcommand{\gbar}{\bar{g}}
 \newcommand{\Kbar}{\bar{K}}
 
 \newcommand{\Obar}{\bar{O}}
 \newcommand{\tbar}{\bar{t}}
 \newcommand{\vbar}{\bar{v}}
 \newcommand{\xbar}{\bar{x}}

 \newcommand{\tinyA}{{\rm\tiny A}}
 \newcommand{\tinyAbar}{{\rm\tiny \Abar}}
 \newcommand{\tinyB}{{\rm\tiny B}}
 \newcommand{\tinymax}{{\rm\tiny max}}
 \newcommand{\tinyL}{{\mbox{\tiny L}}}
 \newcommand{\tinyLT}{{\mbox{\tiny LT}}}

 \newcommand{\arcosh}{{\rm arcosh}}
 
 \newcommand{\artanh}{{\rm artanh}}
 \newcommand{\kom}{{\omega}}
 \newcommand{\komn}{\omega_{\mbox{\tiny 0}}}
 \newcommand{\komlt}{\omega_{\mbox{\tiny LT}}}

\newcommand{\hsp}{\hspace{5mm}}

\newcounter{saveeqn}
\newcommand{\alpheqn}{\setcounter{saveeqn}{\value{equation}}
\stepcounter{saveeqn} \setcounter{equation}{0}
\renewcommand{\theequation}{\mbox{\arabic{saveeqn}-\alph{equation}}}}
\newcommand{\reseteqn}{\setcounter{equation}{\value{saveeqn}}%
\renewcommand{\theequation}{\arabic{equation}}}

\begin{document}

\title{The twin paradox and Mach's principle}

\author{Herbert Lichtenegger\inst{1} \and Lorenzo Iorio\inst{2}
}                     
%
%

\institute{   Institut f\"ur Weltraumforschung, \"Osterreichische Akademie der Wissenschaften,
              Schmiedlstrasse 12, 8042 Graz, Austria \\
              Tel.: +43-4316-425615\\
              Fax: +123-45-678910\\
              \email{herbert.lichtenegger@oeaw.ac.at}           
           \and
             Ministero dell'Istruzione, dell'Universit\`{a} e della Ricerca (M.I.U.R.)-Istruzione \\ International Institute for Theoretical Physics and
Advanced Mathematics Einstein-Galilei\\
Fellow of the Royal Astronomical Society (F.R.A.S.)\\
\email{lorenzo.iorio@libero.it} \\
}

\date{Received: 11 September 2011 / Revised version: 8 November 2011}
%

\abstract{
The problem of absolute motion in the context of the twin paradox is discussed.
It is shown that the various versions of the clock paradox feature some aspects
which Mach might have been appreciated. However, the ultimate cause of the
behavior of the clocks must be attributed to the autonomous status of spacetime,
thereby proving the relational program advocated by Mach as impracticable.
%
} 
\maketitle
\section{Introduction}
\label{intro}
Newtonian physics rests upon the notion of absolute time and absolute space. Free bodies are considered
to be in a state of rectilinear and uniform motion and any deviation from this state is due to a force
acting on the body. Therefore, a force shows up by inducing an acceleration whose value is solely determined
by the inertial mass of the body, while a constant motion is not conceivable in any physical experiment.
Uniform motion is a relative quantity and only reasonable when related to a frame of reference. Thus the
Newtonian concept of mechanics implies the existence of a special class of reference frames (so called
inertial systems) which are characterized by the absence of inertial forces and which differ among each
other only be their constant and uniform state of motion. All physical phenomena proceed in the same way
in all inertial frames and therefore these frames cannot be distinguished from each other by any physical
means (principle of relativity).

Although inertial frames play a fundamental role in the description of physical processes,
the fact that absolute motion is inherent in classical mechanics has been shown already by Newton himself
by means of his famous rotating bucket experiment: Newton was led to the conclusion that the forces responsible
for the curvature of the surface of the water in the rotating bucket cannot be attributed to the relative motion
between the water and the wall of the bucket, but are rather induced by the accelerated motion with respect to
absolute space, thereby giving the notion of absolute space a meaning of its own.

Mach considered Newton's view of absolute space as a meaningless and idle metaphysical concept which
had to be avoided in any empirical theory. Rather, physics should rest upon observable effects and
motion should only be determined with reference to other bodies and not with respect to empty space
\cite{mach:1960}. In particular, inertial forces should be induced by accelerated motion with respect
to all other bodies in the universe. This viewpoint, that an inertial frame of reference and the inertial
mass of a body are determined by the mass distribution in the universe and that the inertial force is
due to the gravitational action of distant matter is the basis of what is usually referred to as Mach's
principle\footnote{The philosophical basis of Mach's principle is rooted in the rejection against the
introduction of any unobservable quantities in the natural sciences.  Hence in scientific investigations
economy is the ultimate ambition, i.e. not more given quantities should be assumed than are absolutely needed.
According to Mach, the starting point of all our perception is the immediate experience and every
proposition must be reducible to sensations. These ''elements of sensation'' constitute the elementary
building blocks of observations and are thus the basis for all scientific disciplines.}
\cite{barbour&pfister:1995,him&mashhoon:2007}. It should be noted, however, that over the years Mach's principle
has been interpreted in numerous ways, and depending on the formulation of this principle, different versions of
it may even lead to opposite conclusions \cite{bondi&samuel:1997,rindler:1994}.

Following Mach, Einstein was also dissatisfied with the obviously preferred class of inertial frames
and the ability of absolute space to act upon matter without being affected in turn by matter \cite{einstein:1956}.
By means of his equivalence principle, i.e. the local indistinguishability between gravitational and inertial
forces, Einstein hoped to surmount this problem by putting down inertial interactions to gravitational ones.

However, although having initially in mind the relativity of all motion, general relativity does not overcome
the problem of absolute motion: the idea of the relativity of motion is in fact incompatible with a theory
that includes local Lorentz invariance and contains  Newtonian gravity in a correspondence limit. It is
essentially the conception of the gravitational field as the curvature of spacetime which opposes the relativity
of arbitrary motion \cite{mashhoon:1988,mashhoon:1994}.

In the following we discuss the problem of absolute motion in the context of the paradox of the relativistic
twins and investigate whether the different aging of the twins conforms to Machian ideas, i.e. whether
the different settings of the clocks can be explained by solely invoking observable phenomena. As will be
shown, the answer will somewhat depend on what aspect is emphasized, although the existence of spacetime
itself ultimately contradicts Mach's principle.

\section{Moving clocks}
\label{sec:1}
The twin paradox will be discussed for the following situations
\begin{list}{}{}
\itemsep-10pt
\item[(a)] clocks in uniform motion\\
\item[(b)] clocks in a compact space\\
\item[(c)] accelerated clock\\
\item[(d)] two clocks in free fall\\
\item[(e)] two clocks counterorbiting a rotating mass.
\end{list}

\subsection{Clocks in uniform motion}
\begin{figure}[h]
\centering
\epsfig{file=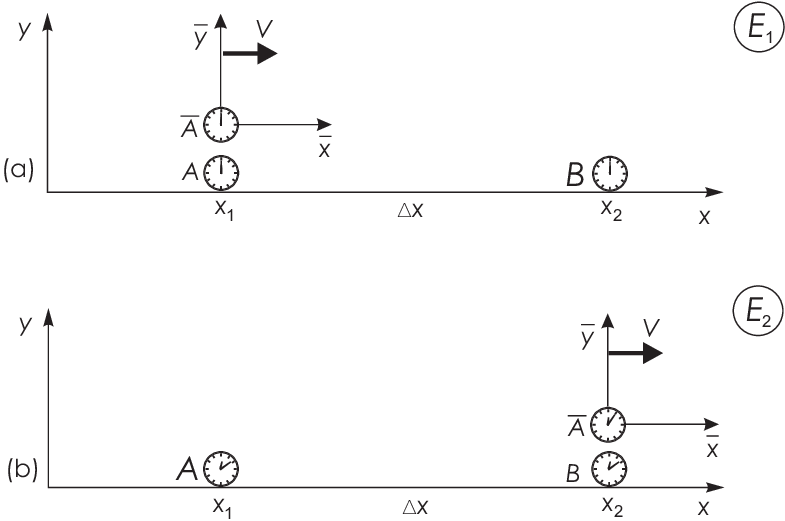,width=9cm,angle=0}
\caption{\small Clock $\Abar$ uniformly moving in the frame $K$ from $A$ to $B$.}
\end{figure}

In the following we discuss a modified version of the standard twin paradox which avoids any acceleration
of the clocks and hence the question whether general relativity is required to resolve the problem.
We consider an inertial system $K$ with two identical clocks $A$ and $B$ at rest separated by the distance
$\Delta x$ and a uniform moving clock $\Abar$ with velocity $V$ (Fig. 1a). At event $E_1$ the locations of
$A$ and $\Abar$ shall coincide and these clocks shall also show the same time both in $K$ and in the rest
frame $\Kbar$ of $\Abar$, i.e. $t_\tinyA(E_1)=t_\tinyAbar(E_1)=\tbar_\tinyA(E_1)=\tbar_\tinyAbar(E_1)$.
Since $A$ and $B$ are at rest in $K$ we can assume them to be synchronized with respect to $K$, i.e
$t_\tinyA=t_\tinyB$ at any time.
Now we ask for the time shown by $B$ and $\Abar$ when they meet each other at event $E_2$ (Fig. 1b).

\paragraph{Observer in \boma{K}}
With respect to $K$, the time $\Delta t$ needed for $\Abar$ to move the distance $\Delta x$ is given by
$\Delta x/V$ and hence the time $t_\tinyB(E_2)$ of $B$ at $E_2$ is
\beq\label{uniform.K.tB}
t_\tinyB(E_2)=t_\tinyB(E_1)+\Delta t=t_\tinyB(E_1)+\frac{\Delta x}{V}=t_\tinyA(E_1)+\frac{\Delta x}{V}.
\eeq
Since $\Abar$ is moving uniformly in $K$, the time $t_\tinyAbar(E_2)$ kept by $\Abar$ at $E_2$ is reduced
by the factor $\gamma^{-1}=(1-V^2/c^2)^{1/2}$ with respect to $\Delta t$
\beq\label{uniform.K.tA}
t_\tinyAbar(E_2)=t_\tinyAbar(E_1)+\frac{\Delta t}{\gamma}=t_\tinyAbar(E_1)+\frac{\Delta x}{\gamma V}.
\eeq
The time difference between $B$ and $\Abar$ upon their encounter in $K$ is therefore (with $t_\tinyA(E_1)=t_\tinyAbar(E_1)$)
\beq\label{uniform.K.tdiff}
t_\tinyB(E_2)-t_\tinyAbar(E_2)=\frac{\Delta x}{V}\left(1-\frac{1}{\gamma}\right)>0\hsp\Rightarrow\hsp t_\tinyB(E_2)>t_\tinyAbar(E_2),
\eeq
showing that clocks at rest in $K$ run faster than moving ones.

\paragraph{Observer in \boma{\Kbar}}
As seen from $\Kbar$, the system $K$ moves likewise with velocity $V$, however, the distance $\Delta x$
between the two clocks is Lorentz contracted, i.e. $\Delta\xbar=\Delta x/\gamma$ and hence $\Abar$ shows
the time $\Delta\tbar$ elapsed between the two events to be
\beq\label{uniform.Kbar.tAbar}
\tbar_\tinyAbar(E_2)=\tbar_\tinyAbar(E_1)+\Delta\tbar=\tbar_\tinyAbar(E_1)+\frac{\Delta x}{\gamma V}.
\eeq
For an observer comoving with $\Abar$, the clocks $A$ and $B$ are not synchronized because of their motion,
rather $B$ runs ahead of $A$ by the factor $\Delta x V/c^2$, i.e.
\beq\label{uniform.Kbar.sync}
\tbar_\tinyB(E_1)=\tbar_\tinyA(E_1)+\frac{\Delta x V}{c^2}.
\eeq
Further, the time displayed by $B$ in $\Kbar$ during the two events is dilated with respect to $\Delta\tbar$
by the factor $\gamma^{-1}$, therefore the time shown in $\Kbar$ by $B$ upon the encounter with $\Abar$ is
\beq\label{uniform.Kbar.tB}
\tbar_\tinyB(E_2)=\tbar_\tinyB(E_1)+\frac{\Delta\tbar}{\gamma}=\tbar_\tinyA(E_1)+\Delta x\left(\frac{V}{c^2}+
 \frac{1}{\gamma^2 V}\right)=\tbar_\tinyA(E_1)+\frac{\Delta x}{V}
\eeq
and the time difference between $\Abar$ and $B$ reads in compliance with (\ref{uniform.K.tdiff}) (with
$\tbar_\tinyAbar(E_1)=\tbar_\tinyB(E_1)$)
\beq\label{uniform.Kbar.tdiff}
\tbar_\tinyB(E_2)-\tbar_\tinyAbar(E_2)=\frac{\Delta x}{V}\left(1-\frac{1}{\gamma}\right).
\eeq
Although $\Abar$ is at rest in $\Kbar$, it reports less time when consecutively contrasted with a pair of clocks
synchronized in $K$. Finally, comparison of ({\ref{uniform.K.tA}) with ({\ref{uniform.Kbar.tAbar}) and
({\ref{uniform.K.tB}) with ({\ref{uniform.Kbar.tB}) shows that both observers agree with the setting of the
clocks when checked at the same time at the same place. \textrm{Eq.} ({\ref{uniform.K.tA}) \textrm{is the standard time dilation
of a uniformly moving clock and can be found in almost all text books on special relativity.}

\subsection{Clocks in a compact space}
As a model for a spatially closed 2-dimensional spacetime we can visualize a surface of a cylinder which is
constructed from a flat stripe of width $L$ by identifying points at $x=0$ (event $E_0$) with those at $x=L$
(event $E_\tinyL$) \textit{at the same time} (Fig. 2a) \cite{brans&stewart:1973}.
We denote the system, where these identifications are made by $K$ and all events located along a circle around
the cylinder perpendicular to the time axis in $K$ will occur simultaneously and clocks at rest can be synchronized $\rm\grave{a}$ la Einstein
all over the cylinder universe. In an inertial frame $\Kbar$ moving with velocity $V$ with respect to $K$ into
the $\pm x$-direction (the origins of $K$ and $\Kbar$ shall coincide for $t=\tbar=0$), the events to be identified
are spatially and temporarily separated by (Fig. 2a)
\beq\label{compact.synch.gap}
\Delta\xbar=\gamma L,\hsp\Delta\tbar^\pm=\mp\gamma\frac{VL}{c^2},
\eeq
where the $\pm$ superscript indicates the the time shift in $\Kbar$ for its motion in the $+x$- and $-x$-direction,
respectively. Due to this identification, clocks in $\Kbar$ can only be synchronized as long as their spatial distance
is less than $\gamma L$. When trying to synchronize clocks around the entire universe, there will inevitably be somewhere
two adjacent clocks which exhibit a time shift according to (\ref{compact.synch.gap}) in such a way that for the motion
of $\Kbar$ into the $\pm x$-direction the clock being synchronized by a light beam in $\Kbar$ in the $-\xbar$-direction
will be slow/ahead by an amount $\Delta\tbar^\pm$ with respect to a neighboring clock being synchronized by a beam in the
$+\xbar$-direction. As a consequence, bodies circling the cylinder in opposite directions in $\Kbar$ will need different
times for completing one revolution \cite{peters:1983}.

Due to the compactification, two clocks $A$ and $\Abar$ (with $A$ at rest in $K$ and $\Abar$ at rest in $\Kbar$), which meet
at some point (event $E_1$) will encounter again at some other point (event $E_2$), while both clocks remain inertial during
their separation (Fig. 2b). Although inertial observers in $K$ and $\Kbar$ will each see the other clock moving uniformly,
the two clocks will be out of phase at their re-encounter, as discussed in the following.

\begin{figure}[t]
\centering
\epsfig{file=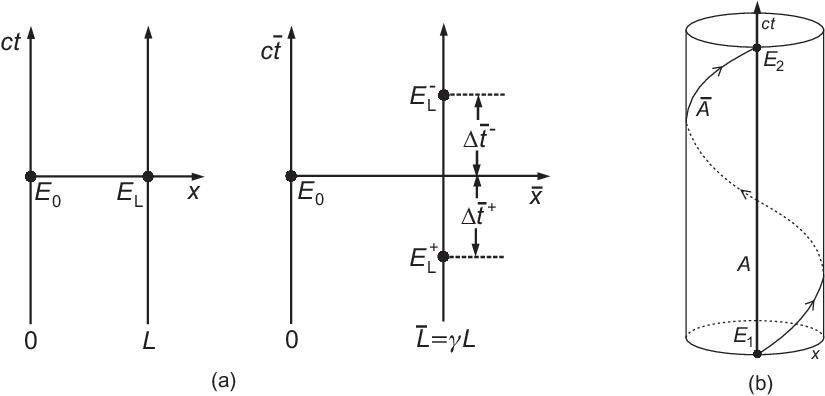,width=9cm,angle=0}
\caption{\small (a) Identification of events in $K$ and $\Kbar$ when $\Kbar$ is moving in the $\pm x$-direction with velocity $V$.
(b) Word lines of $A$ and $\Abar$ in $K$ between two encounters.}
\end{figure}

\paragraph{Observer in \boma{K}}
With respect to $K$, the clock $\Abar$ is moving with constant velocity $V$ and the distance covered by $\Abar$ between two successive
encounters is $L$. The time $\Delta t_\tinyA$ needed to move this distance is $L/V$ and hence the time registered by $A$ at event $E_2$ is
\beq\label{compact.K.tA}
t_\tinyA(E_2)=t_\tinyA(E_1)+\Delta t_\tinyA=t_\tinyA(E_1)+\frac{L}{V}
\eeq
while $\Abar$ displays the reduced time
\beq\label{compact.K.tAbar}
t_{\tinyAbar}(E_2)=t_{\tinyAbar}(E_1)+\frac{\Delta t_A}{\gamma}=t_{\tinyAbar}(E_1)+\frac{L}{\gamma V}.
\eeq
The time difference as recorded by an observer in $K$ is therefore ($t_\tinyA(E_1)=t_\tinyAbar(E_1)$)
\beq\label{compact.K.tdiff}
t_\tinyA(E_2)-t_{\tinyAbar}(E_2)=\frac{L}{V}\left(1-\frac{1}{\gamma}\right)>0,
\eeq
indicating that an observer in $K$ ages faster than one in $\Kbar$.

\paragraph{Observer in \boma{\Kbar}}
For an observer in $\Kbar$, the clock $A$ moves with velocity $V$ a distance $\gamma L$ between the reunion of
the clocks in the time $\Delta\tbar_\tinyAbar=\gamma L/V$. In addition, $\Abar$ will display the synchronization
gap (\ref{compact.synch.gap}) present in $\Kbar$. Hence, $\Abar$ will show in $\Kbar$ at event $E_2$ the time
\beqa\label{compact.Kbar.tAbar}
\tbar_\tinyAbar(E_2)&=&\tbar_\tinyAbar(E_1)+\Delta\tbar_\tinyAbar\pm\Delta\tbar^\pm=
 \tbar_{\Abar}(E_1)+\frac{\gamma L}{V}-\gamma\frac{VL}{c^2}\nonumber\\
&=&
 \tbar_{\tinyAbar}(E_1)+\frac{\gamma L}{V}\underbrace{\left(1-\frac{V^2}{c^2}\right)}_{1/\gamma^2}=
 \tbar_{\Abar}(E_1)+\frac{L}{\gamma V},
\eeqa
where the $\pm$ sign corresponds to the motion of $\Kbar$ into the $\pm x$-direction.
Further, for an observer in $\Kbar$ the clock $A$ will not register the time shift (\ref{compact.synch.gap}),
therefore $A$ is expected in $\Kbar$ to record the time
\beqa\label{compact.Kbar.tA}
\tbar_\tinyA(E_2)&=&\tbar_\tinyA(E_1)+\frac{1}{\gamma}\Bigl[\tbar_\tinyAbar(E_2)-\tbar_\tinyAbar(E_1)\mp\Delta\tbar^\pm\Bigr]=
 \tbar_\tinyA(E_1)+\frac{1}{\gamma}\left(\frac{L}{\gamma V}+\gamma\frac{LV}{c^2}\right)\nonumber\\
&=&
 \tbar_\tinyA(E_1)+\frac{L}{V}\underbrace{\left(\frac{1}{\gamma^2}+\frac{V^2}{c^2}\right)}_{1}=\tbar_\tinyA(E_1)+\frac{L}{V}.
\eeqa
Again both observers agree upon the setting of their clocks during the encounter and upon the time difference
(\ref{compact.K.tdiff}) shown by the clocks. It should be noted that this time difference is identical with
expressions (\ref{uniform.K.tdiff}) and (\ref{uniform.Kbar.tdiff}) obtained in case (a) for two clocks separated
by a distance $\Delta x=L$. \textrm{Further discussions of the twin paradox in a cylindrical universe can be found in e.g.
\cite{low:1990,peters:1983} and the relations (\ref{compact.K.tA})-(\ref{compact.K.tAbar}) are derived in e.g.
\cite{barrow&levin:2001,brans&stewart:1973,dray:1990}}.

\subsection{Clocks in non-uniform motion}
\begin{figure}[h]\label{fig3}
\centering
\epsfig{file=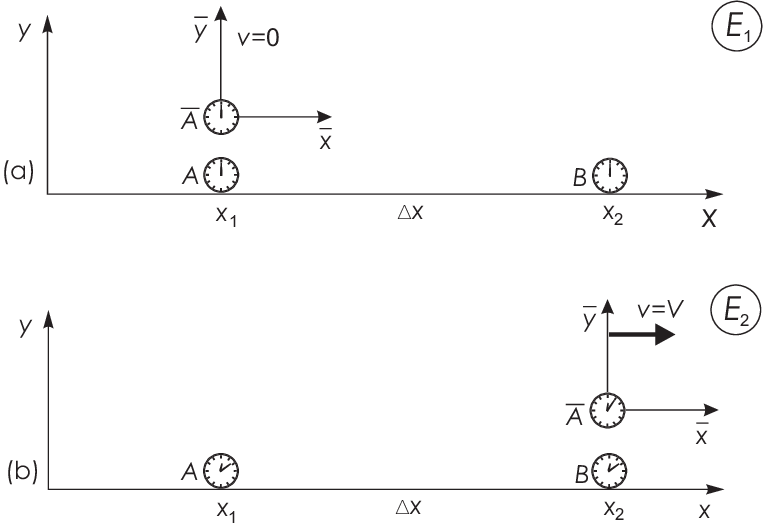,width=9cm,angle=0}
\caption{\small Clock $\Abar$ accelerated with constant proper acceleration from $A$ to $B$, thereby
acquiring the velocity $V$.}
\end{figure}

\textrm{Clocks in hyperbolic motion are discussed in a number of papers, e.g.
\cite{eriksen&gron:1990,iorio:2005,jones&wanex:2006,minguzzi:2005,nikolic:2000,styer:2007}.
In the conventional clock paradox, of two clocks $A$ and $\Abar$ originally at rest in an inertial system, $\Abar$ is accelerated and departs
from $A$. Later, upon deceleration, it eventually comes at rest, reverses its motion and returns back to $A$ in a similar way. As in case (a)
it suffices to compare the readout of the accelerated clock with that of two synchronized clocks at rest.} Therefore, we
consider the inertial system $K$ with the clocks $A$ and $B$ at rest, synchronized (i.e. $t_\tinyA=t_\tinyB$)
and separated by some distance $\Delta x$.
Now a third clock $\Abar$ close to $A$ is initially likewise at rest and all three clocks shall show the same
time $t_1=t(E_1)=\tbar(E_1)$ (see Fig. 3a). At event $E_1$, $\Abar$ is uniformly accelerated with respect to its rest system
$\Kbar$ into the direction of $B$ and takes up the velocity
$V$ when passing over $B$ (event $E_2$, see Fig. 3b). Again we ask for the time indicated by $B$ and $\Abar$ when they
meet each other at event $E_2$.

\paragraph{Observer in \boma{K}}
Since the acceleration $\gbar$ of $\Abar$ is constant in $\Kbar$, the acceleration of $\Abar$ in relation to $K$
is given by means of the Lorentz transformations\footnote{The use of the Lorentz transformation between accelerated
systems is based on the hypothesis of locality: at each instant of time the momentary rest frame of an accelerated
observer is assumed to be physically equivalent with an inertial frame with the same instantaneous velocity
\cite{mashhoon:2003}.} via \cite{rindler:1991}
\beq\label{g(v)}
\frac{dv}{dt}=\frac{\gbar}{\gamma^3(v)}
\eeq
and integration yields the time in $K$ needed for $\Abar$ to attain the velocity $v$
\beq\label{t(v)}
t-t_1=\gamma(v)\frac{v}{\gbar}.
\eeq
Because $\Abar$ has velocity $V$ when passing by $B$, the latter clock will indicate the time
\beq\label{nonuniform.K.tB}
t_\tinyB(E_2)=t_\tinyB(E_1)+\gamma(V)\frac{V}{\gbar}
\eeq
during the event $E_2$. Further, from (\ref{t(v)}) the velocity of $\Abar$ in $K$ at any time $t$ is found to be
\beq\label{v(t)}
v_\tinyAbar(t)=\frac{\gbar(t-t_1)}{\sqrt{1+[\gbar^2(t-t_1)^2]/c^2}}.
\eeq
The time dilatation of $\Abar$ accumulated in $K$ between the events $E_1$ and $E_2$ can be calculated via
\beq
t_\tinyAbar(E_2)-t_\tinyAbar(E_1)=\int\limits_{t(E_1)}^{t(E_2)}\frac{dt}{\gamma[v_\tinyAbar(t)]}
\eeq
and together with (\ref{t(v)}) and (\ref{v(t)}) the time displayed by $\Abar$ at $E_2$ is given by
\beq\label{nonuniform.K.tAbar}
t_\tinyAbar(E_2)=t_\tinyAbar(E_1)+\frac{c}{\gbar}\artanh\frac{V}{c}.
\eeq

\paragraph{Observer in \boma{\Kbar}}
An observer comoving with $\Kbar$ notices a static pseudo-gravita\-tional field and sees the clock $B$ falling freely
towards $\Abar$ with an acceleration \cite{moller:1972}
\beq\label{nonuniform.Kbar.d2xdt2}
\frac{d^2\xbar}{d\tbar^2}=\frac{2\gbar}{c^2(1+\gbar\xbar/c^2)}\left(\frac{d\xbar}{d\tbar}\right)^2-
 \gbar\left(1+\frac{\gbar\xbar}{c^2}\right).
\eeq
From this equation, the velocity and position of $B$ at any time $\tbar$ is found to be
\beq\label{nonuniform.Kbar.vbar}
\vbar_\tinyB(\tbar)=-c\left[1+\frac{\gbar\xbar_\tinyB(E_1)}{c^2}\right]
 \frac{\sinh\gbar[\tbar-\tbar_\tinyAbar(E_1)]/c}{\cosh^2\gbar[\tbar-\tbar_\tinyAbar(E_1)]/c}
\eeq
and
\beq\label{nonuniform.Kbar.xbar}
\xbar_\tinyB(\tbar)=\frac{c^2}{\gbar}\left[\left(1+\frac{\gbar\xbar_\tinyB(E_1)}{c^2}\right)
 \frac{1}{\cosh\gbar[\tbar-\tbar_\tinyAbar(E_1)]/c}-1\right].
\eeq
Now the time needed in $\Kbar$ for $B$ to move the distance $|\xbar_\tinyB(\tbar)-\xbar_\tinyB(E_1)|$ follows
immediately from (\ref{nonuniform.Kbar.xbar})
\beq\label{nonuniform.Kbar.tbar}
\tbar-\tbar_\tinyAbar(E_1)=\frac{c}{\gbar}\arcosh\frac{1+\gbar\xbar_\tinyB(E_1)/c^2}{1+\gbar\xbar_\tinyB(\tbar)/c^2}.
\eeq
Upon setting $\xbar_\tinyB(\tbar)=0$ and substituting (\ref{nonuniform.Kbar.tbar}) into (\ref{nonuniform.Kbar.vbar}),
the velocity of $B$ when passing the origin of $\Kbar$ is given by
\beq
\vbar_\tinyB(E_2)=V=c\tanh\frac{\gbar[\tbar_\tinyAbar(E_2)-\tbar_\tinyAbar(E_1)]}{c}
\eeq
and hence the arrival time of $B$ at $\Abar$ finally reads
\beq\label{nonuniform.Kbar.tAbar}
\tbar_\tinyAbar(E_2)=\tbar_\tinyAbar(E_1)+\frac{c}{\gbar}\artanh\frac{V}{c}
\eeq
in accordance with (\ref{nonuniform.K.tAbar}). \textrm{The result (\ref{nonuniform.Kbar.tAbar}) can also be found in some
textbooks (e.g. \cite{moller:1972,rindler:1991}) and papers (e.g. \cite{eriksen&gron:1990,iorio:2005}).}
The time difference shown by the clocks $B$ and $\Abar$ at event $E_2$ is thus ($t_\tinyB(E_1)=\tbar_\tinyAbar(E_1)$)
\beq\label{nonuniform.Kbar.tdiff}
t_\tinyB(E_2)-\tbar_\tinyAbar(E_2)=\gamma(V)\frac{V}{\gbar}\left[1-\frac{c}{\gamma(V)V}\artanh\frac{V}{c}\right]>1
\eeq
again indicating that the moving clock in $K$ lags behind those at rest.


\subsection{Clocks in the vicinity of massive objects}
\begin{figure}[t]\label{fig4}
\centering
\epsfig{file=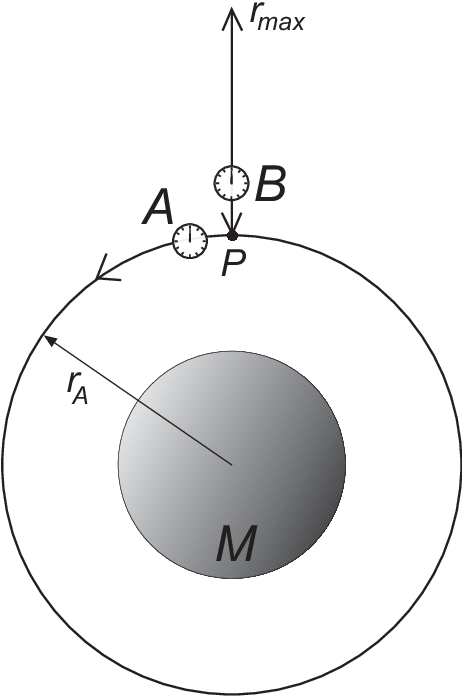,width=5cm,angle=0}
\caption{Two clocks $A$ and $B$ in the vicinity of a massive object $M$. $A$ is moving along a circular orbit with
radius $r_\tinyA$, $B$ along a purely radial orbit with the turning point at a distance $r_\tinymax$.}
\end{figure}
Let us consider a massive object of mass $M$ and radius $R$ and two clocks $A$ and $B$ initially located at $P$
(event $E_1$).
While $A$ is assumed to follow a circular orbit with radius $r_\tinyA$, $B$ shall be on a radial trajectory
with a turning point at the distance $r_\tinymax$ (see Fig. 4). The radial velocity of $B$ is chosen in such a way that
$A$ and $B$ will meet again at $P$ after one revolution of $A$ (event $E_2$). As in the previous sections we
will determine the proper time elapsed by $A$ and $B$ when they meet each other again at event $E_2$.

We denote the spacetime coordinates of $E_1$ and $E_2$ with $(r_1,\,t_1)$ and $(r_2,\,t_2)$, respectively,
where $t$ is the coordinate time. The motion of $B$ must be suitably adjusted in order to achieve the requirement
of $r_1=r_2=r_\tinyA$ at $E_2$. The spacetime geometry around $M$ is given by the Schwarzschild metric and implies
that the coordinate time period $T_t^\tinyA$ for one revolution of $A$ about $M$ is identical with the Kepler
period in Newtonian gravity
\beq\label{massive.tA}
t_2^\tinyA-t_1^\tinyA\equiv T_t^\tinyA=2\pi\sqrt{\frac{r^3_\tinyA}{GM}},
\eeq
while the proper time $\tau$ for one period reads
\beq\label{massive.tauA}
\tau_2^\tinyA-\tau_1^\tinyA\equiv T_\tau^\tinyA=2\pi\sqrt{\frac{r^3_\tinyA}{GM}\left(1-\frac{3}{2}\frac{R_s}{r_\tinyA}\right)}=
 T_t^\tinyA\sqrt{1-\frac{3}{2}\frac{R_s}{r_\tinyA}}
\eeq
with $R_s=2GM/c^2$ being the Schwarzschild radius.

In the Schwarzschild field the equation for the radial motion is similar to its expression in the Kepler case
\beq\label{massive.rp}
\frac{dr}{d\tau}=-\sqrt{2GM\left(\frac{1}{r}-\frac{1}{r_\tinymax}\right)}
\eeq
and it is convenient to introduce an angular parameter $\theta$ which is connected with the
distance $r$ and the proper time $\tau$ via the cycloidal relation \cite{misner_et_al:1973}
\alpheqn
\beqa
r&=&\frac{r_\tinymax}{2}(1+\cos\theta)\label{massive.cycloit-a}\\
\tau&=&\sqrt{\frac{r_\tinymax^3}{8GM}}\,(\theta+\sin\theta).\label{massive.cycloit-b}
\eeqa\reseteqn

The coordinate time needed to fall freely from $r_\tinymax$ to $r_\tinyA$ is given by \cite{misner_et_al:1973}
\beqa\label{massive.t}
t&=&
 \frac{1}{c}\left\{R_s\ln\frac{Q+\sqrt{r_\tinymax/r_\tinyA-1}}{Q-\sqrt{r_\tinymax/r_\tinyA-1}}+\right.\nonumber\\
&&\left.+
 \frac{Qr_\tinymax}{2}\left[\left(1+\frac{2R_s}{r_\tinymax}\right)\arccos\left(\frac{2r_\tinyA}{r_\tinymax}-1\right)+
 \sqrt{1-\left(\frac{2r_\tinyA}{r_\tinymax}-1\right)^2}\right]\right\}\nonumber\\
Q&=&\sqrt{\frac{r_\tinymax}{R_s}-1},
\eeqa
from which $r_\tinymax$ can be determined upon putting $t=T_t^\tinyA/2$. If we denote the value of $\theta$
satisfying (\ref{massive.cycloit-a}) (with $r=r_\tinyA$) by $\theta^*$, the proper time $\tau_\tinyB(E_2)$
displayed by $B$ at the event $E_2$ is found by inserting $\theta^*$ into (\ref{massive.cycloit-b}), while
$\tau_\tinyA(E_2)$ is
obtained with the help of (\ref{massive.tauA}). Hence, the ratio of the proper times of $B$
and $A$ at their re-encounter reads
\beq
\frac{\tau_\tinyB(E_2)}{\tau_\tinyA(E_2)}=
 \frac{\theta^*+\sin\theta^*}{\pi\sqrt{(1+\cos\theta^*)^3\left(1-\frac{3}{2}\frac{R_s}{r_\tinyA}\right)}}>1.
 \label{massive.ratio}
\eeq
Fig. 5 illustrates the ratio $\tau_\tinyB/\tau_\tinyA$ as a function of $r_\tinyA/R_s$, showing that $A$ runs slower
than $B$ and hence displays less time than $B$ at event $E_2$. \textrm{This complies with the result of \cite{holstein&swift:1972}
who studied the relative aging of two observers orbiting a central mass, one along a circular orbit and the other one
along an eccentric orbit. They concluded that the proper time is shorter for the observer
who travels the longer space path, which is also the case in our example since the up and down path of $B$ is shorter
than the perimeter covered by $A$. Recently, \cite{ambramovicz&bajtlik:2009} considered the situation of two twins in the vicinity
of a massive body: one twin is circling about the central mass while the other twin stays fixed at some point on the circle by
applying an appropriate radial acceleration. As noted by \cite{ambramovicz&bajtlik:2009} the non-accelerated twin (i.e. the one circling around)
ages slower than the accelerated one, in contrast to the standard twin paradox (case(c)), where the accelerated twin (i.e. the one leaving
the earth and coming back) is younger. However, in the case described by \cite{ambramovicz&bajtlik:2009} it is again the twin traveling
the longer space path who records less proper time. On the other hand, \cite{gron&braeck:2011} presented an example which appears to
be opposite: in the case of one twin staying fixed in the vicinity of a massive object and the other twin moving radially upward and
downward, it is the latter twin who is older upon reunion.}
\begin{figure}[t]
\centering
\epsfig{file=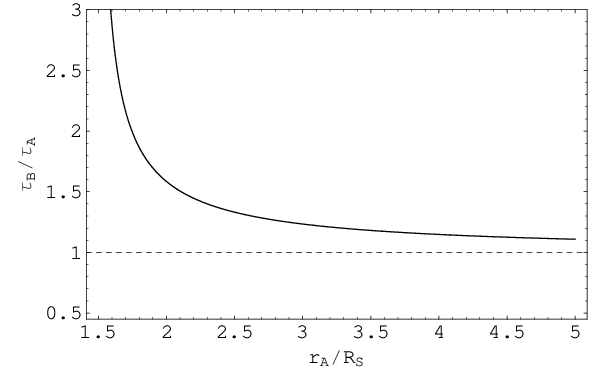,width=9cm,angle=0}
\caption{\small Ratio of proper times as a function of the radius of the circular orbit. Note that
$\tau_\tinyB/\tau_\tinyA$ tends to infinity for $r_\tinyA/R_s\rightarrow 3/2$.}
\end{figure}

\subsection{Clocks in orbits around a massive rotating object}
\begin{figure}[b]
\centering
\epsfig{file=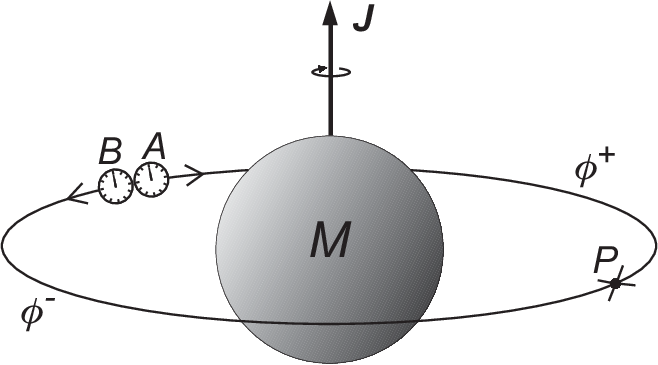,width=8cm,angle=0}
\caption{\small Clocks along opposite circular orbits around a rotating body with mass $M$ and angular momentum $J$. The first
meeting point after the separation of the clocks is denoted by $P$.}
\end{figure}
We consider a rotating mass $M$ with spin $J$ and two clocks on identical circular orbits with radius $r$ moving in
opposite directions in the plane perpendicular to $J$ (Fig. 6). Let us denote the clocks along the pro- and retrograde
orbit by $A$ and $B$, respectively, and calculate the proper times $\tau_\tinyA$ and $\tau_\tinyB$ that passes between
two successive meetings of the clocks. If the geometry around $M$ is described by the linearized Kerr metric,
then the equation of motion for the azimuthal coordinate $\phi$ yields
\beq
\bigg|\frac{d\phi}{dt}\bigg|_\pm\equiv\kom_\pm=\komn\pm\komlt,
\eeq
where
\beq
\komn=\sqrt{\frac{GM}{r^3}},\hsp\kom_\tinyLT=-\komn^2\frac{J}{c^2M}
\eeq
are the Kepler- and Lense-Thirring frequency, respectively, and the plus and minus sign corresponds
to the pro- and retrograde direction. Since
\beq
\phi_++\phi_-=2\pi,\hsp \phi_\pm=\kom_\pm t,\hsp t=\frac{2\pi}{\kom_++\kom_-}=\frac{\pi}{\komn},
\eeq
where $t$ is the coordinate time elapsed between two successive encounters of the clocks, we find
by means of
\beqn
\frac{d\tau}{dt}=1-\frac{3}{2}\frac{GM}{c^2r}-\frac{9}{8}\frac{(GM)^2}{c^4r^2}\pm \frac{3GJ}{c^4r}\komn
\eeqn
the difference in proper time shown by the clocks $A$ and $B$ at their first meeting point \cite{him_et_al:2000,markley:1973,tartaglia:2000}
\beq
\tau_\tinyA-\tau_\tinyB=6\pi\frac{GJ}{c^4 r}.
\eeq
Therefore, an observer along a prograde orbit around the central mass will age faster than
a counterrotating observer.

\section{Machian or anti-Machian?}
Common to all five examples is the fact that all observers agree upon the differential aging of the clocks,
both in sign and in magnitude. At first glance, the different position of the watch-hands in case (a) at the
encounter of the clocks might
be surprising since only a constant relative motion is involved. However, special relativity is based on
an absolute quantity, namely the speed of light which has the same value in all inertial frames and which
entails the relativity of simultaneity. The synchronization in $K$ can thus be performed by sending light signals
to the two clocks from a point halfway in between their positions (Einstein synchronization) and this procedure
is totally intrinsic to $K$ without reference to any external system. Therefore, inertial frames can not be
distinguished by this method\footnote{The Einstein synchronization instruction is only one among many possible
synchronization procedures. Other choices, e.g., based on the reference of external systems can destroy the
symmetry among the inertial systems and may not lead to the relativity of simultaneity. However, though the
procedures may be different, the observational consequences will be similar to those implied by the Einstein
synchronization, showing the conventional character of the synchronization process
\cite{erlichson:1985,minguzzi:2002,selleri:2005}.}.
The asymmetry in (a) is thus induced by the introduction of a preferred frame in which $A$ and $B$ are
synchronous and by comparing the single clock $\Abar$ with these clocks at different locations in $K$.
The consideration of two synchronized clocks $\Abar$ and $\Bbar$ in $\Kbar$ and a single moving clock $A$
would reveal the opposite result, i.e. $A$ would show less time now than $\Bbar$ when passing over it. In
addition, the situation in the two reference frames is not symmetrical: while an observer comoving with
$\Kbar$ will see two clocks in motion ($A$ and $B$) and one at rest ($\Abar$), an observer in $K$ will see
two clocks at rest ($A$ and $B$) and one in motion ($\Abar$). In this sense, although the two synchronized
clocks will single out their rest frame, the outcome of the measurements of uniformly moving clocks could
be considered Machian, since no relationship to any unobservable entity is established and the time
difference depends only on the relative velocity between the clocks.
On the other hand, it is difficult to conceive that a relative velocity alone could represent a physical
mechanism responsible for the different reading of the clocks. Therefore, since no masses are involved
in the process of the time measurements which could produce a dynamical effect that might be liable for
the asymmetric result, it is only Minkowski spacetime itself which can cause the clocks to run differently.
This, however, is not in line with Machian ideas.

The flat cylindrical spacetime (b) is constructed in such a way that for an observer $O$ whose world line is
parallel to the axis of the cylinder, two simultaneously emitted light rays will also arrive simultaneously
at the opposite points of the cylinder. This gives $O$ a privileged status: he is the only observer in the
cylinder universe who can synchronize his clocks \textit{all over} spacetime and who measures the smallest
circumference $L$ of the universe. Any observer $\Obar$ moving with respect to $O$ and trying to synchronize
clocks in his rest frame beyond a distance $\gamma L$ will fail to do so since the specific topology imposed
by $O$ prevents his entire spacetime to be covered by a single coordinate patch. Moreover, the perimeter of
the universe as determined by $\Obar$ is always larger than the one detected by the preferred observer $O$.
These features of a compact space certainly contradict the spirit of Mach, because they grant spacetime an
autonomy which cannot be traced back to any observational cause.

In case (c) of the accelerated clock it may be noted that it is not essential that $\Abar$ is compared
with different clocks at different locations since $A$ and $B$ represent a pair of synchronized clocks.
Alternatively, one could equally well consider $\Abar$ being first accelerated to the velocity $V$ and
subsequently being decelerated again to rest in $K$. A similar but opposite movement would then bring
$\Abar$ back to $A$ and it would still lack behind $A$\footnote{Because all four hyperbolic acceleration
phases are identical except for the sign, $\Abar$ would keep a time four times larger than the value
given by (\ref{nonuniform.K.tAbar}) or (\ref{nonuniform.Kbar.tAbar}).} \cite{iorio:2005}. The asymmetry
in the lapse of the proper times is due to the privileged status of inertial frames being unmodified
in special relativity.

It might be tempting to put the cause for the differential aging down to the relative acceleration between
the two clocks, viewing either $A$ and $B$ or $\Abar$ as the ''distant masses''. However, the time difference
depends on neither of the masses, rather it depends on the maximum velocity reached and the constant proper
acceleration of the ''truly'' accelerated clock $\Abar$. Moreover, there is no unambiguous relative
acceleration, since the acceleration in $K$ becomes smaller with increasing time while it remains constant
in time for $\Abar$. It should also be noted that the metric associated with (\ref{nonuniform.Kbar.d2xdt2})
is given by
\beq
g_{00}=\left(1+\frac{\gbar\xbar}{c^2}\right)^2,\hsp g_{ij}=-\delta_{ij}
\eeq
which implies a vanishing Riemann curvature tensor. Hence the gravitational field in $\Kbar$ cannot be generated
by real masses in the universe as opposed to what Mach might have been expected.

In case (d) and (e) the state of affairs becomes different because now real masses are also comprised in the
''elements of sensation''. Moreover, the situation may appear less obvious with regard to the previous examples since
the two clocks are freely falling between their encounters and therefore observers comoving with the clocks
experience identical local physical laws.
The difference in their proper time can thus be not a consequence of any local effects, rather it must be
attributed to some global influence. In fact the observers could trace back their differential aging to
their different motion with respect to the central mass, showing that there are multiple geodesics connecting
the same events or, in other words, that their geodesics are differently embedded within the global spacetime
manifold. Another evidence for the observers to be in an asymmetric position in a gravitational field is due
to the different tidal forces existing in their local reference frames \cite{durso_et_al:1973}.
Hence the behavior of the clocks can be attributed to the existence of the central (spinning)
mass; indeed in the formal limit $M\rightarrow 0$ (or $J\rightarrow 0$) the two clocks would tick in consonance.
Interpreted in this way, i.e. that the different tick rates can be ascribed to the gravitational influence of
massive bodies, clocks moving in a gravitational field may even reveal some Machian features.

\section{Conclusion}

Even though some properties of the various clock paradoxa may be considered as Machian, the ultimate cause for
the behavior of the clocks is based on the absolute character of spacetime, since it is always the clock with
the shorter world line which runs ahead. Although the specific form of the metric will be determined by the
mass-energy content of spacetime, its very existence is still independent of it.
In order to conform with the Machian idea of the relativity of all motion, spacetime should loose its metric
properties in a universe void of all mass-energy. However, this is excluded by the relativity theory since
the components of the metric tensor at best reduce to their Minkowskian values in the limit of an empty space.
It is precisely this ontological aspect, that spacetime -- besides specifying the relations between bodies --
is granted a substance in its own, which makes the relational program unfeasible that Mach might have been envisioned.

\section*{Acknowledgements}
L.I. thanks B. Mashhoon for inspiring correspondence on the topic of this paper long ago.
%
%
%

%
%

\end{document}